# Recording and manipulation of vagus nerve electrical activity in chronically instrumented unanesthetized near term fetal sheep

Aude Castel[1]*, Burns, Patrick M[1]*, Benito, Javier[1], Liu, Hai L[2], Kuthiala, Shikha[2], Durosier, Lucien D[2], Frank, Yael S[3], Cao, Mingju[2], Marilène Paquet[4], Fecteau, Gilles[1], Desrochers, André[1], Martin G. Frasch[2,3,5]

[1] Department of Clinical Sciences, Faculty of Veterinary Medicine, University of Montreal, St-Hyacinthe, QC, Canada

[2] Department of Obstetrics and Gynaecology and Department of Neurosciences, CHU Ste-Justine Research Centre, University of Montreal, Montreal, QC, Canada

[3] Department of Obstetrics and Gynecology and Center on Human Development and Disability (CHDD), University of Washington, Seattle, WA, USA

[4] Département de pathologie et de microbiologie, Faculté de Médecine Vétérinaire, Université de Montréal, Saint-Hyacinthe (Québec)

[5] Centre de recherche en reproduction animale (CRRA), University of Montreal, St-Hyacinthe, QC, Canada

**\* equal contribution**

**Corresponding author:**
Martin G. Frasch
Department of Obstetrics and Gynecology
University of Washington
1959 NE Pacific St
Box 356460
Seattle, WA 98195
Phone: +1-206-543-5892
Fax: +1-206-543-3915
Email: mfrasch@uw.edu

**Short Abstract:**
Little is known about the vagus nerve activity in near-term fetuses. The chronically instrumented unanesthetized fetal sheep model is used to study human fetal physiology because it permits chronic instrumentation with catheters and electrodes, which allow repetitive blood sampling, substance injection and recording of bioelectrical activity. We describe the procedures required to manipulate the vagus nerve activity in this model.

**Long abstract:**
**Background:** The chronically instrumented pregnant sheep has been used as a model of human fetal development and responses to pathophysiologic stimuli. This is due to the unique amenability of the unanesthetized fetal sheep to the surgical placement and maintenance of catheters and electrodes, allowing repetitive blood sampling, substance injection, recording of bioelectrical activity, application of electric stimulation and in vivo organ imaging. Recently, there has been growing interest in pleiotropic effects of vagus nerve stimulation (VNS) on various organ systems such as innate immunity, metabolism, and appetite control. There is no approach to study this in utero and corresponding physiological understanding is scarce.
**New Method:** Based on our previous presentation of a stable chronically instrumented unanesthetized fetal sheep model, here we describe the surgical instrumentation procedure allowing successful implantation of a cervical uni- or bilateral VNS probe with or without vagotomy.
**Results:** In a cohort of 53 animals, we present the changes in blood gas, metabolic, and inflammatory markers during the postoperative period. We detail the design of a VNS probe which also allows recording from the nerve. We also present an example of vagus electroneurogram (VENG) recorded from the VNS probe and an analytical approach to the data.
**Comparison with Existing Methods:** This method represents the first implementation of VENG/VNS in a large pregnant mammalian organism.
**Conclusions:** This study describes a new surgical procedure allowing to record and manipulate chronically the vagus nerve activity in an animal model of human pregnancy.

**Introduction**

Vagal nerve stimulation (VNS) has been long used for the treatment of drug-resistant epilepsy.(Morris and Mueller, 1999) More recently, this clinically well-tolerated treatment approach has been explored in multiple animal experimental models and clinical trials for the treatment of a number of conditions involving the brain's control of the innate immune system.(Kwan et al., 2016) The underlying substrate for this approach is the cholinergic anti-inflammatory pathway (CAP), mediated by the vagal nerves.(Pavlov and Tracey, 2017) Much remains to be studied about manipulating the vagal nerve activity and the physiology of its pleiotropic effects. The emerging field of bioelectronic medicine has tackled this challenge. In the field of fetal development, chronically instrumented pregnant sheep have been used extensively for 50 years as a highly translational animal model.(Burns et al., 2015; Durosier et al., 2015; Gotsch et al., 2007; Herry et al., 2016; Liu et al., 2016; Morrison et al., 2018; Nitsos et al., 2006; Svedin et al., 2005; Yan et al., 2004)

The extensive use of the fetal sheep model is due to the unique amenability of the unanesthetized fetal sheep to the surgical placement and maintenance of catheters and electrodes, allowing repetitive blood sampling, recording of bioelectrical activity, application of electric stimulation and *in vivo* brain imaging.(Burns et al., 2015; Carmel et al., 2012; Cortes et al., 2017) Here we describe a novel methodology of vagal nerve manipulation in near-term fetal sheep which allows the study of fetal vagus nerve activity in relation to other biophysical and biochemical parameters. A brief version of the approach, without representative results, appeared recently in a book chapter format.(Frasch et al., 2018a) The present manuscript delineates the methodology in greatest detail providing representative physiological findings and outlining possible pitfalls and future applications. This instrumentation technique can be useful to further elucidate the development and function of neuroimmunological circuits such as the CAP.

**Material and Methods**

Animal care followed the guidelines of the Canadian Council on Animal Care and the approval by the Université de Montréal Council on Animal Care (protocol #10-Rech-1560).
Animal selection: Healthy pregnant ewes were selected. The ewes were barrier nursed at all times to minimize the enzootic potential such as *Coxiella burnetii*. Gloves and masks (N-95 type) were worn during any interaction with the animal including by non-surgical personnel.

1. **Anesthesia of the ewe**

A single-lumen catheter is inserted into a jugular vein of the ewe. The ewe is sedated using acepromazine (Atravet 10 mg/mL) 2 mg intravenously approximately 30 minutes prior to the induction of anesthesia in order to reduce the stress associated with the procedure preventing an increase in levels of cortisol. Induction of general anesthesia is performed using diazepam (Diazepam 5 mg/mL) 20 mg, ketamine (Ketalar 100 mg/mL) 4-5 mg/kg and propofol (Propofol 10 mg/mL) 0.5 to 1 mg/kg intravenously (IV).
An airway exchange catheter is inserted into the trachea using a laryngoscope with a Wisconsin type blade (Extra-long 350 mm Left-Handed Blade) to aid with the intubation. A silicone endotracheal tube (9 to 12mm Inner Diameter) is slid off the airway exchange catheter and into the trachea. This technique aids visualization of the vocal cords during the intubation process. The cuff of the endotracheal tube is inflated carefully to avoid pressure-induced ulceration of the trachea and the tube is attached to the head of the ewe with a sling.

The endotracheal tube is connected to the respiratory circuit of the anesthetic machine and mechanical ventilation is begun immediately. The ventilator settings are adjusted to maintain a $P_aCO_2$ within normal limits of 35 to 45 mmHg.
A (22 to 20 G; 1 in [0.9 x 25 mm] to 1.16 in [1.1 x 30 mm]) catheter is inserted into the auricular artery and connected to non-compliant tubing to monitor direct arterial blood pressure.

A multi-parameter physiologic monitor can be used to record intraoperatively the electrocardiogram (ECG), direct arterial blood pressure, oxygen saturation ($SpO_2$), capnography ($P_{ET}CO_2$), and temperature every five minutes. All physiologic data are transferred via a serial cable to a central physiologic data-collecting computer. Normal body temperature (38.5-39.9ºC or 101-103.9ºF) is maintained using a circulating water blanket. Fluid therapy in the form of a balanced poly-ionic solution is administered at 10 mL/kg for the first hour of general anesthesia and then reduced to 5 mL/kg/h.

Trimethoprim-sulfadoxine (5 mg/kg IV to the ewe just prior to the skin incision as antibiotic prophylaxis. Standard aseptic techniques are used for all surgical manipulations of the ewe and the fetus.

**2. Surgical procedure**
**2.1 Overview of the surgical procedure**

An incision is made through the lower abdominal wall immediately cranial to the udder through the linea alba while carefully avoiding the median mammary vein.
The greater omentum is retracted cranially and the uterine horns are then palpated manually from the body of the uterus to the tip of the horn to determine the number of fetuses and their size. If there is more than one fetus, the larger fetus is chosen by evaluating manually the head size and the width between the orbits. The head of the chosen fetus is held firmly through the partially exteriorized uterus and a hysterotomy is performed on the large curvature. A non-latex sterile surgical glove filled with sterile saline is immediately placed over the head of the fetus. Alternatively, moistened gauze can be used to keep the fetal head moist. The uterus is secured to the abdominal wall using Babcock forceps.

The thoracic limbs are exteriorized and the fetus is gently pulled out of the uterus up to the xiphoid process. Polyvinyl catheters are inserted into the right and left brachial vein and arteries using a standard cut-down technique. Another polyvinyl catheter is inserted into the amniotic cavity with its end fixated to the sternum of the fetus. Each catheter is attached to a needle connected to a double stopcock to allow for subsequent blood sampling or pressure monitoring.

Stainless steel electrodes are sutured to the manubrium, xiphoid process and to each point of the shoulder to monitor the electrocardiogram (ECG).

Bilateral cervical VNS probes are installed and followed by vagotomy (Vx) distal or proximal to the VNS probes. The probe design is shown in Figure 1. We also deposited a video of the procedure in the FigShare repository.(P. M. Burns et al., 2020)

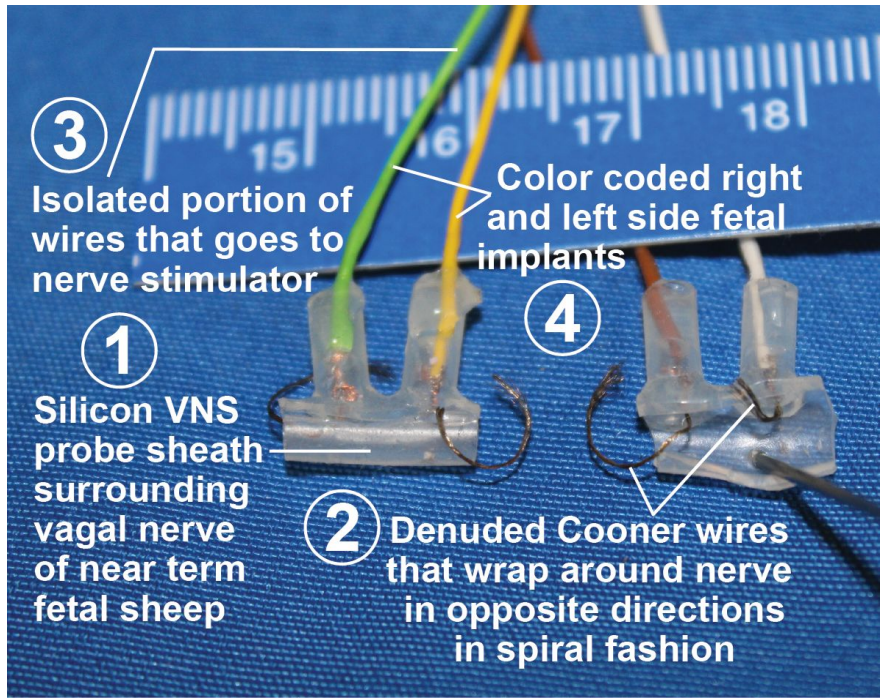

**Figure 1. Design of the vagus nerve stimulation probe**. A silicon tube of 3.17 mm inner diameter and 6.4 mm outer diameter (Micro Medical Tubing made by Scientific Commodities Inc, Cat# BB31775-V/15) was slit open to be placed as an insulating sheath surrounding the vagus nerve. Two silver Cooner wires are denuded on the inside of the silicon tube to be wrapped around the nerve in opposite directions from each other in a spiral fashion. The isolated portion of these wires connects to the nerve stimulator. The isolated portion is color-coded to help distinguish left and right side implants once the fetus is returned to the mother's womb. This external portion of the wires is made by Metrofunk Kabel GmbH. Depending on the nerve diameter, a smaller tubing can be used such as 1.68mm inner diameter x 2.39mm outer diameter (Micro Medical Tubing made by Scientific Commodities Inc, Cat# BB31785-V/11).

Clenbuterol (30 µg per sheep) is slowly administered intravenously to the ewe over 15 minutes prior to returning the fetus to the uterus to prevent uterine spasm due to surgical manipulation and reduce post-operative discomfort both of which should reduce the stress of the ewe and fetus.

The fetus is returned to the uterus and a small stab incision is made in the left flank for all catheters and electrodes to exit.

The hysterotomy incision is closed using a three-layer closure. The abdominal wall and the subcutaneous space are closed with a synthetic absorbable suture. The skin incision is closed with surgical staples.

**2.2. A detailed description of the surgical procedure**

2.2.1. Surgical preparation

The wool is removed by shaving from the xiphoid process to the mammary gland and along the fold of the flank on each side with a blade #40. Then, the ventral abdomen is cleaned thoroughly with a 4% chlorhexidine gluconate with detergent on a soft brush for 3 minutes.
A standard sterile scrub with chlorhexidine gluconate 4% with detergent is performed starting from the center of the abdomen and progressing in a centrifugal fashion for 3 minutes. Sterile irrigation saline is poured on the abdomen to remove the disinfectant soap. For the final step of the surgical preparation, three alternate passages of chlorhexidine gluconate solution 2% and isopropyl alcohol 70% are performed.

2.2.2. Hysterotomy

Adequate depth of anesthesia is ensured prior to incision. A standard 20 cm midline laparotomy incision is made from the umbilicus to just cranial to the udder through the linea alba to minimize abdominal muscle damage. Prior to skin incision, the median mammary vein is located and carefully retracted while incising the midline. The skin is incised with a scalpel blade #21 while the linea alba is incised with a blade #10.

A long sponge forceps is inserted in the abdomen along the left abdominal wall up to the planned exit site for catheters in the paracostal region. The tips of the forceps are pressed against the wall for an assistant to locate it and confirm the appropriate site. The jaws of the forceps are opened slightly (1cm) and a 2 cm full-thickness stab incision is made by the assistant with a #21 scalpel blade.

The tips of the forceps are exteriorized through the incision, opened again, to gently grasp the catheters and pull them out of the abdomen through the ventral abdominal incision.

The uterus is palpated to determine the fetal position, number, and size. To identify the largest fetus, the inter-aural distance is compared. The uterine wall is incised with Metzenbaum scissors over 10 cm on the large curvature and over the dorsum of the fetal head, avoiding the cotyledons.

A blunt-ended cannula (teat cannula designed for dairy cattle) is inserted through the placental membranes to obtain an amniotic fluid sample free from hemorrhage. Then, the placental membranes are incised using scissors. The cranial half of the fetus is exteriorized through this incision and a sterile non-latex surgical glove filled with sterile saline at 37 °C is placed over the fetus's head to help maintain normothermia. During removal of the fetal upper body from the uterus, Babcock forceps holding the placental membranes and uterine wall are held up by an assistant to prevent the loss of amniotic fluid. Then, using 6 Babcock forceps, the fetal membranes and uterine wall are clamped to the skin to prevent abdominal contamination with amniotic fluid.

2.2.3. Fetal instrumentation

Only the parts of the fetal body that need to be instrumented are exposed while the remaining parts are kept inside the uterus or covered by moist and warm (37 °C) sterile cloths.

The fetus' thoracic limbs are abducted to facilitate exposure to the brachial artery and vein bilaterally. An incision is made along the medial aspect of both antebrachium and the tissue around the brachial artery and vein is carefully dissected.

All catheters used had been sterilized by gas at an approved facility. Polyvinyl catheters pre-marked with black lines every 2 cm are inserted into the right and left brachial arteries and the left brachial vein using a standard cut-down technique. For this purpose, the vessel to be catheterized is freed from adjacent tissue over 1 cm and a ligature is placed at the distal portion of the vessel with a braided synthetic absorbable USP 2-0 suture. A ligature is also preplaced at the proximal aspect of the vessel but kept untied.

Using Castroviejo scissors, the vessel is cut transversally to approximately 30% of its diameter. The blood flow is stopped partially by pulling on the proximal suture and the catheter is inserted in a proximal direction up to 8 cm or until resistance is detected and then pulled back slightly. The proximal aspect of the catheter is temporarily secured using a vascular clamp. Another suture is placed around both the proximal and distal aspect of the catheter while an assistant is continuously aspirating and flushing the catheter to ensure its patency. The fetal skin is closed in a continuous suture pattern using a USP 2-0 braided synthetic absorbable suture.

Insulated stainless steel electrodes are attached to the right and left shoulder, manubrium and xiphoid process to allow for monitoring of the fetal ECG. The amniotic pressure and sampling catheter is sutured to the sternum. This catheter is fenestrated at its extremity.

All the catheters are secured on the dorsum of the fetus using USP 2-0 braided synthetic absorbable suture material.

Before the fetus is placed back into the uterus, clenbuterol 30 µg IV is administered slowly over 15 minutes to the ewe to avoid hypotension and ensure uterine relaxation while closing it.

2.2.4. Fetal vagal nerve instrumentation and vagotomy

To install bilateral cervical VNS probes, the fetus head is extended and the neck is kept straight and stable while approaching the vagal nerves. A 3 cm skin incision is made adjacent to the jugular vein. The tissues are bluntly dissected with Kelly forceps until the carotid is located with the vagus nerve adjacent to it. Two centimeters of the nerve are freed from the carotid with Mosquito forceps. A double T shape silicone sleeve attached to electrodes is gently slid around the vagal nerve with each electrode circling the nerve. The silicone sleeve is then secured using USP 3-0 braided synthetic absorbable suture material.

Bilateral cervical Vx is performed distal or proximal to the VNS probes. Prior to Vx, the silicone sleeve and its electrodes are slid away from the incision site. The vagal nerve is held by an assistant with a tissue forceps while the nerve is cut. The extremity of the nerve with the VNS probes is secured to the silicone sleeve with USP 4-0 braided synthetic absorbable suture material. The sleeve is then attached to the surrounding muscles. The skin is closed using USP 2-0 braided synthetic absorbable suture material in a continuous pattern.

2.2.5. Closure and post-operative care

The fetal membranes are closed using USP 4-0 braided synthetic absorbable suture material in a continuous pattern. Only one catheter or electrode are incorporated into the closure at a time to ensure tightness. The uterine muscular layer is closed in a double layer Cushing pattern respecting Halsted principles with USP 0 braided synthetic absorbable suture material. The surgical knots are carefully buried.

Using a purse-string suture pattern, all the catheters and ECG cables are secured as they exit the left paracostal incision.

Using a USP 1 monofilament synthetic absorbable suture material, the linea alba is closed with a continuous pattern. The subcutaneous tissues are also closed in a continuous pattern using USP 2-0 braided synthetic absorbable suture material. The skin layer is closed with surgical stainless steel staples.

Ampicillin (250 mg) is administered intravenously to the fetus as well as via the amniotic catheter into the amniotic cavity. Lost amniotic fluid is replaced with warm saline.

All the exteriorized catheters and ECG electrodes are placed into a bag to maintain sterility. A stockinette around the torso of the ewe is used to secure all the catheters and electrodes.

General anesthesia is terminated, and the ewe is extubated once laryngeal reflexes have returned to normal.

The ewe is returned to a metabolic cage once stable enough following general anesthesia and will reside in it for the duration of the experiment. The cage should allow the ewe to stand, lie down and eat *ad libitum* while monitoring the unanesthetized fetus without sedating the mother. Oxygen is provided to the ewe via an intranasal cannula inserted into one nostril and sutured in place.

Buprenorphine (300μg IV) is administered every 6 to 8 hours during the immediate postoperative period. For the following three days, antibiotics are administered prophylactically to the ewe (Trimethoprim sulfadoxine 5 mg/kg IV) and the fetus (250 mg of sodium ampicillin intravenously and again via the amniotic catheter).

The metabolic status of both ewe and fetus are evaluated with blood gas analyses.

All catheters are flushed slowly once a day with approximately 5 mL of saline per line after antibiotic prophylaxis. using the smallest volume of heparinized saline possible with caution not to exceed the daily dose of heparin and fluids permissible to the fetus in order to prevent fluid overload.

### 3. Data recording and analysis

During surgery, the maternal and fetal ECG and heart rates as well as maternal arterial blood pressure and airway pressure (Paw) can be recorded continuously.(Burns et al., 2015) All maternal data except ECG are acquired with Life Window Monitor from DIGICare Biomedical Technology and these data are fed into a Micro 1401 analog-digital converter along with fetal and maternal ECG signals. Maternal and fetal ECG are passed first into the 1902 amplifier. All data are then recorded and displayed in Spike 2 software (CED, Cambridge, UK).

A 1 mL arterial sample is simultaneously taken from the ewe and fetus for arterial blood gas analysis, lactate, glucose, and base excess determination (ABL-820, Radiometer) at the beginning of the fetal surgery (immediately after inserting the first arterial catheter) and after closing the uterus.

During postoperative recovery, a 3 mL fetal blood sample is collected to measure the IL-6 and TNF-α inflammatory profiles. The plasma is centrifuged at 4 °C (4 min, 4000 x g), decanted and stored at -80 °C for subsequent ELISA testing.

For the purpose of the reported representative results, the animals were sacrificed using intravenous injection of 20 mL sodium pentobarbital six days after surgery. Fetal growth was assessed by evaluating the body, brain, liver and maternal weights. The duration of the experimental period may vary depending on the design chosen for a given research goal and can reach up to 6 weeks.

### 4. Cytokine analyses (optional step)

Plasmatic cytokine concentrations (IL-6, TNF-α) are determined by using an ovine-specific sandwich ELISA. Mouse anti-sheep monoclonal antibodies (to capture antibody against IL-6) or mouse anti-bovine monoclonal antibody (for TNF-α) are used at a concentration of 4 μg/mL as pre-coat on ELISA plate incubated at 4 °C overnight after a 3 times wash with washing buffer (0.05% Tween 20 in PBS, PBST).

The plates are then blocked for 1 h with 1% BSA in PBST followed by a 3 times wash with a washing buffer.

Recombinant sheep proteins (IL-6, TNF-α) are used as ELISA standard and serial dilutions are prepared to range from standard 1 of 2000 ng/mL to standard 7 of 31.25 pg/mL.

Fifty µL of serially diluted protein standards and samples are loaded in each well and incubated for 2 hours at room temperature. The plates are washed again 3 times. All standards and samples were run in duplicate.

Fifty µL of rabbit anti-sheep polyclonal antibodies (to detect antibody IL-6) or rabbit anti-bovine polyclonal antibody (to detect TNF-α) at a dilution of 1:250 are added to each well and incubated for 30 min at room temperature. Following this step, the plates are washed 5 times with a washing buffer.

Fifty µL of the goat anti-rabbit IgG-HRP conjugated are added to the wells (dilution 1:5000) for 30 minutes.

Incubation is performed with 50 µL of TMB substrate solution per well.

Color development reaction is stopped at the desired time with 25 µL of 2 N sulphuric acid.

The plates are read using an ELISA plate reader at 450 nm, with a 570 nm wavelength correction.

In this study's assays, the sensitivity for IL-6 ELISA was 16 pg/mL and the sensitivity for TNF-α ELISA was 13.9 pg/mL. For all assays, the intra-assay and inter-assay coefficients of variance were <5% and <10%, respectively.

**5. Vagus nerve histological analysis** (optional step)

After the animal had been euthanized, the fetal left and right cervical vagus nerves were retrieved from within the VNS probes and each vagus nerve segments (around 2 cm in length) were collected . To prevent curling of tissues and to keep the cranio-caudal orientation, each individual vagus nerve was pinned to a small piece of cardboard and fixed in 10% neutral buffered formalin at room temperature for 24 h.

After fixation, the nerves were placed in a histology cassette and oriented in between sponges.

The samples were processed, embedded in paraffin, cut and stained with H&E according to general histology procedures.

H&E stained slides were examined by a board-certified veterinary pathologist and graded for inflammation and axonal degeneration.

**6. Vagus nerve electroneurogram (VENG).**

Fetal heart rate (FHR), ECG, and arterial blood pressure were monitored continuously (1902 amplifier and micro3 1401 ADC by CED, Cambridge, U.K., and NL108A, NeuroLog, Digitimer, Hertfordshire, U.K) and sampled at 256 and 1000 Hz, respectively. VNS was applied via

NeuroLog's NL512/NL800A using pulse sequence pre-programmed in Spike 2. The VNS settings were as follows: DC rectangular 5 V, 100 uA, 2 ms, 1 Hz according to (Borovikova et al., 2000). VENG was recorded at 10,000 Hz sampling rate. Based on literature and our experience in the piglet model established subsequently to this model(P. Burns et al., 2020), a sampling rate of 20,000 Hz may be more appropriate.

## 7. Statistical analyses

General linear models (GLM) were used to test for any significant differences in the blood gases and in histomorphological scoring of the vagus nerve (SPSS). The VENG analysis was conducted with the open source EEGLAB package v2019_1 within the Matlab environment (but can also be run in the open source Octave) (Matlab 2013b for Linux, MathWorks, Natick, MA). Mutual information was computed using the open source library Java Information Dynamics Toolkit (JIDT) in Python 3.6.8.(Lizier, 2014)

The data is presented as mean and standard deviation (SD) with a p-value of less than 0.05 being considered significant.

**Results**

The main objective of the presented approach is to demonstrate the feasibility of surgical manipulation of the ovine fetal cervical vagus nerves by Vx and/or VNS with VENG. The combination of Vx and VNS approaches allows for a specific efferent (distal of Vx) or afferent (proximal of Vx) VNS for mechanistic dissection of peripheral and central nervous influences of vagus nerve signaling.

In the following, we review representative findings for blood gas and metabolites, systemic inflammation, FHR response to vagotomy, vagus nerve histology after VNS/VENG probe implantation, and VENG recordings patterns in this animal model. As an important part of the model, we present an in-house low-cost design of the VNS probe which we tested for both VNS and VENG (Fig. 1).

Fifty three pregnant time-dated ewes were instrumented at 126 days of gestation (dGA, ~0.87 gestation, term 145 dGA) with arterial, venous, and amniotic catheters and electrocardiogram (ECG) electrodes with sterile technique as described above under general anesthesia. The total duration of the procedure was about 2 h. Not all reported measurements were performed in every animal.

*Fetal arterial blood gas and acid-base status*

Fetal physiological characteristics during surgery, vagotomy, and VNS probe instrumentation were reported in Table 1 of (Herry et al., 2019) and were within the physiological norm for the gestational age of the fetus. Vx had a measurable but not drastic effect on the fetal arterial blood gas and acid-base status when comparing measurements at surgery start (prior to Vx), surgery end (after Vx), and at 72h postoperatively. Specifically, pH, $O_2$Sat, and BE showed similar recovery dynamics in both groups but were higher in Vx than sham animals. That is, Vx animals were slightly less acidotic than the sham animals.

*Effect of Vx on systemic inflammation*

Fetal arterial IL-6 ELISA was performed in 27 control and 26 vagotomized animals. The assay rendered values below the sensitivity threshold of 16 pg/mL throughout the post-operative recovery period in 93% of all measurements. There was no difference in IL-6 values on postoperative days one to four between the vagotomized animals and controls.
On postoperative day one, Vx and controls measured 1±2 and 3±11 pg/mL, respectively (p=0.23); on postoperative day two, the values were 1±2 and 2±5 pg/mL, respectively (p=0.46); on postoperative day three, we measured 2±6 and 2±6 pg/mL (p=0.97); on day four the values were 1±4 and 1±3 pg/mL, respectively (p=0.83).

Similarly, the TNF-α levels also remained unchanged and very low at 29 pg/mL with ~30% of the animals also showing values below the sensitivity threshold of 13.9 pg/mL throughout the postoperative recovery period.

*Impact of Vx/VNS probe on vagus neuropathology*

We studied whether attaching a VNS probe (shown in Fig. 1) resulted in a higher rate of nerve lesions in the area of the attachment compared to the nerves where no VNS probe was attached. Six pairs of cervical vagus nerves without VNS probe were studied (control group, 3 pairs were shams and 3 pairs were subjected to Vx) and ten pairs with VNS probe installed proximally (= afferent VNS, n=7 pairs) or distally (= efferent VNS, n=3 pairs) to the Vx. An average of 17±5 mm of vagus nerve tissue was examined histologically.

The following lesions were studied and quantified on a scale from 0 to 4, 0 being absent and 4 being > 51% of the specimen. Grade 1 corresponded to modest (-10%), Grade 2= mild(11-30%), Grade 3= moderate(31-50%), Grade 4= severe(+51%) lesion of the respective kind as listed below:

- Degenerative changes (Wallerian degeneration, ischemic changes)(affecting the distal half segment).
- Degenerative changes (Wallerian degeneration, ischemic changes)(affecting the proximal half segment).
- Multifocal granulocytic inflammation with interstitial edema and hemorrhages in the proximal segment.

Overall, nerves subjected to either Vx or Vx/VNS probe installation exhibited modest to mild signs of degenerative or inflammatory lesions compared to sham controls, as would be expected (Table 1). We observed no axonal swelling across the groups. In one pair of nerves collected from the control group, we observed a mild to moderate focal granulocytic inflammation with interstitial edema and hemorrhages. We found a modest focal perivascular lymphocytic infiltrate in one nerve of the sham control pair.

We found a significant main effect of group on the presence of lesions in GLM analysis ($p<0.001$) which was explained by the afferent and efferent VNS groups (both $p=0.014$), but not by Vx+LPS800 alone ($p=0.269$). The group effect on distal degeneration was also significant at $p<0.001$, explained by the efferent VNS group ($p<0.001$) which is consistent with the installation of the VNS probe distal from Vx. The effect of group on proximal degeneration was also significant at $p<0.001$ and explained by the afferent VNS group ($p<0.001$), again consistent with the installation of the VNS probe proximal to Vx. There was no significant effect of the group on the occurrence of the inflammation (multifocal granulocytosis) ($p=0.749$).

*Effects of VNS and Vx on FHR*

Acute VNS showed a known left-right asymmetry with regard to its impact on FHR as shown in Figure 2A: left VNS had no effect, while right VNS reduced the heart rate transiently by ~10%. The right Vx resulted in a rebound transient FHR increase to ~5% above the baseline, exacerbated by further 5% after the right Vx and normalizing after that. Of note, the subsequent left Vx had no impact on this FHR normalization. Performing afferent VNS after Vx had no impact on FHR (Fig. 2B).

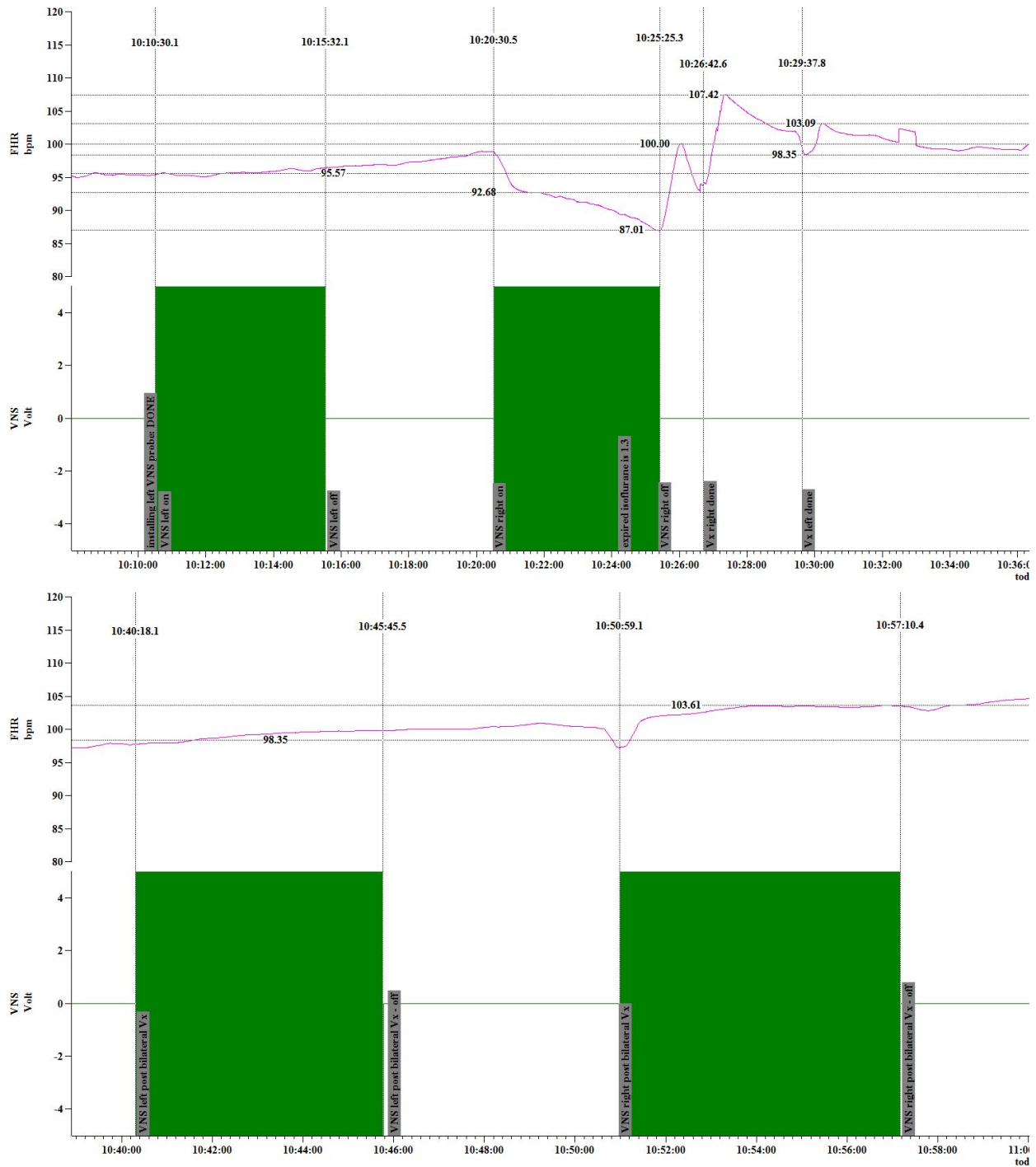

**Figure 2. Effect of VNS on FHR during surgery** prior to vagotomy (Vx) and after Vx revealing a left-right asymmetry (**A**). Afferent VNS post Vx has no effect on FHR regardless of the side of the vagus nerve stimulated (**B**).
FHR, fetal heart rate in beats per minute (bpm); VNS, vagus nerve stimulation (V). The X-axis is the time scale in the hh:mm:ss.

Over the course of the entire surgery and the following 72 hours, as we reported elsewhere, Vx did not result in a change of FHR.(Herry et al., 2019)

*Representative findings of VENG studies*

The qualitative and quantitative differences in VENG sampled at 10,000 Hz and recorded using our VNS/VENG electrode from the right and left vagus nerves were evaluated. The effects of Vx on VENG were also assessed by recording efferent VENG signals following Vx.

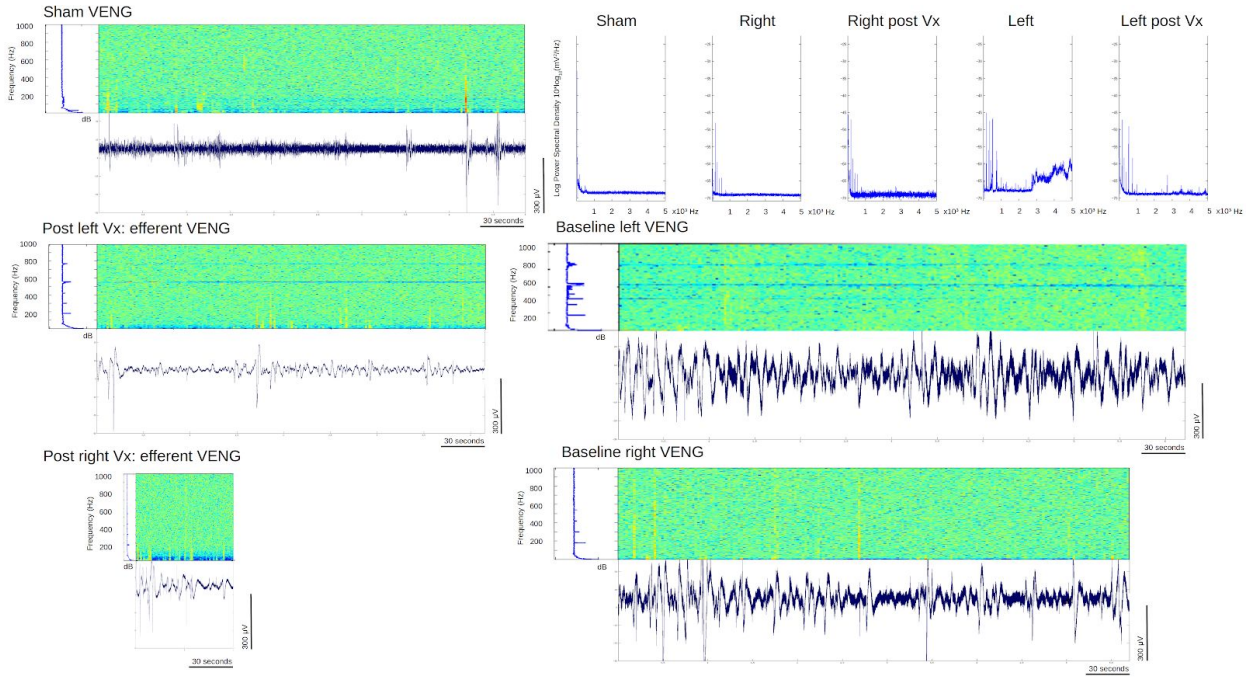

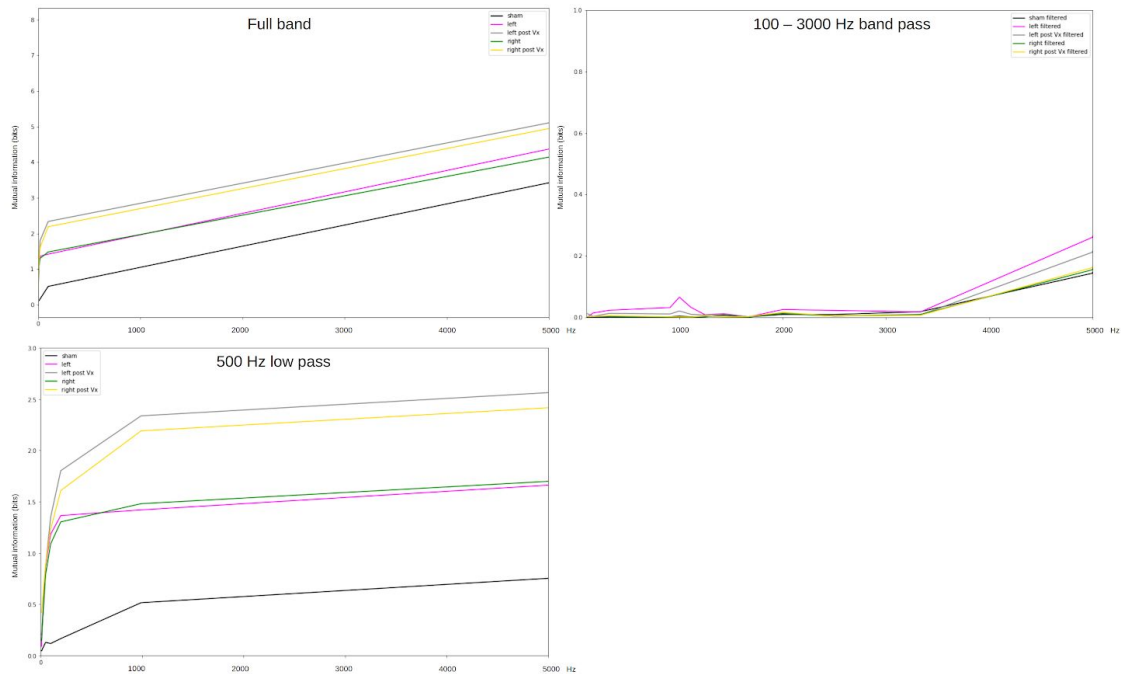

**Figure 3.** The figure presents a VENG analysis pipeline and representative results series using animal 669 (afferent VNS intra-Sx VENG recording), available at figshare. VENG signal could be recorded from one vagus nerve at a time. The sequence of recordings was as follows: sham baseline (no VNS/VENG electrode attached), right VENG baseline followed by vagotomy, followed by post-vagotomy recording, followed by recording from the left VENG, left vagotomy and recording after a vagotomy. High resolution versions of all images and additional graphics are provided in the Supplement on FigShare.(P. M. Burns et al., 2020)

A. Raw VENG, VENG power spectrum/wavelet transform. Each power spectrum's Y axis is optimized for better visualization, so not comparable between the cases. For comparison, see Fig. 3B. Each segment duration is ~5 min except post right VENG (1 min, shorter due to artifacts). The range of VENG is in µV: +/- 120-200 µV.
   TOP RIGHT: Group comparison of power spectra for each measurement. Additional power spectra figures are provided in the Supplement on FigShare.(P. M. Burns et al., 2020)
B. Changes in mutual information function reveal specific peaks at low (< 150 Hz) and high (> 800 Hz) frequencies.
   Black, sham VENG; pink, left VENG; grey, left post Vx VENG; green, right VENG; yellow, right post Vx VENG. X axis, frequencies in Hz, from 0 to 5000; Y axis, mutual information in bits.

Figure 3A shows differences in each recording for VENG signal amplitude, power spectrum, and overtime as a periodogram using wavelet analysis (EEGLAB, Matlab for Linux, MathWorks). At 10,000 Hz sampling rate combined with our relatively simple VNS/VENG electrode attached around the nerve sheath, it is to be expected that considerable noise will be captured. Still, one can readily tell the difference in amplitude and power spectrum distributions between sham VENG and the nerve recordings as well as the effect of vagotomy. The presence of higher frequencies, above 500 Hz, is consistent with earlier reports.(Patel and Butera, 2015; Rozman and Ribaric, 2007)

The high frequency content is also reflected in the peaks of mutual information decays shown in Figure 3B. For easiness of interpretation, we expressed the time scales of the decay of information flow expressed by the auto-mutual information function in Hz to correspond directly to power spectra shown in Figure 3A. The information content generally decreases, the further one goes in the future from a given point in time.(Frasch et al., 2009; Martin G. Frasch et al., 2007; M. G. Frasch et al., 2007) This translates into larger values at higher frequencies, i.e., shorter time scales. Specifically, across all time scales it is overall apparent that the sham VENG shows the least amount of information (black line). There are clear differences in the mutual information between left vagus, right vagus as well pre and post Vx recordings. These differences are frequency-specific with local maxima of information distinct from the sham VENG pattern. We used a 500 Hz low pass filter or 100 - 3000 Hz bandpass filter (EEGLAB) to further accentuate the mutual information differences at lower and higher frequencies. A peak appears at ~1.7 Hz in left and right VENGs, but not in the sham recording; another peak is seen at 100 Hz in both post-Vx recordings, but not seen in left VENG or seen at a lower amplitude in right VENG (Fig. 3B, 500 Hz low pass). Another prominent peak is revealed at 1000 Hz in left VENG which decreases after Vx and is not seen in the right or sham VENG. These insights suggest a relatively simple approach to identifying signaling patterns in these complex recordings that can be associated with other physiological observations.

**Discussion**

The goal of this report is to describe an approach to implement techniques from the emerging field of bioelectronic medicine in the setting of fetal neuroscience. For this purpose, the well-established pregnant sheep model was used as it allows chronic instrumentation of the fetal sheep and unanesthetized monitoring of its physiological activities. The generic approach to animal instrumentation and the method of studying cholinergic signaling on the molecular level have been presented elsewhere. (Burns et al., 2015; Cortes et al., 2017) Of note, the latter aspect represents a direct molecular counterpart to the systems-level approach to study cholinergic signaling described in the present work.(Frasch et al., 2018a) Together, these cellular and integrative approaches comprise a multi-scale paradigm to gain a better understanding of cholinergic signaling, an evolutionary highly preserved mode of communication.(Andersson and Tracey, 2012; Faltine-Gonzalez and Layden, 2018; Frasch et al., 2018b)

Together, the anesthetic and surgical procedures presented on the chronically instrumented unanesthetized fetal sheep with bilateral cervical vagus nerve stimulation and recording allow to establish an animal model for studying fetal vagus nerve physiology and pathophysiology. The entire instrumentation procedure is described in detail for the sake of completeness and to ensure that the entire procedure can be reproduced seamlessly using the presented methodology. Five critical steps during the surgical procedure should be emphasized. First, when passing the catheters and electrodes through the maternal flank, it is important for this step to be done at once to avoid any internal organ injuries. Second, securing the uterotomy operating site prior to exteriorizing the fetus is crucial to prevent or minimize loss of amniotic fluid and subsequent suturing of the amniotic membrane prior to uterine closure. Third, in regards to arterial catheterization, fetal sheep arteries are about 1-2 mm in diameter, therefore, a team of two surgeons will perform better and quicker in this task which saves time and minimizes the overall length of the procedure. Fourth, the steps for bilateral dissection and installation of VNS probes in the fetal neck can be performed before or after the steps for catheter placements, depending on the surgeon's preference. Fifth, careful securing and organization of all catheters and electrodes in the amniotic cavity prior to returning the fetus into the amnion and closing the uterus helps to avoid accidental pulling of the catheters or tearing of the ECG and VNS/VENG electrodes due to maternal or fetal movements after surgery.

A key-point of the study is the description of a method to make in-house cost-effective VNS probes, implant them in the fetus as well as stimulate and/or record vagus nerve electrical activity acutely during surgery or chronically for days. The advantage of the presented VNS probe design is its simplicity allowing the recording of the summated VENG of the entire fiber bundles comprising the vagus nerve. It is also a drawback, as the detailed fiber-specific nerve activity cannot be captured in this manner. More sophisticated electrode designs are technically feasible, at a much higher cost, and will allow fiber-specific micrometer-scale insertion and data acquisition. This also remains a logical future next step in the evolution of such models and does not need to be limited to sheep. Other large mammals known for their relevance in perinatal biology and medicine research, such as piglets, are also great candidates for the recording of VNS/VENG.(P. Burns et al., 2020)

The vagus nerve influences brain function and body metabolism in a pleiotropic manner.(Pavlov and Tracey, 2015, 2012) A new field of bioelectronic medicine has been emerging with the aim to devise therapeutic approaches using VNS to modulate the endogenous salutary signaling of the vagus nerve.(Borovikova et al., 2000; Kwan et al., 2016; Pavlov and Tracey, 2017) For example, VNS has been shown to reduce sympathetic tone, stress-induced anxiety behaviors, and depression symptoms in animal models and in clinical studies.(Caliskan and Albrecht, 2013; Clancy et al., 2014; George et al., 2008; Lamb et al., 2017; Liu et al., 2013; O'Keane et al., 2005; Pena et al., 2014; Ylikoski et al., 2017)

VNS is thought to facilitate tonic inhibition of the basolateral amygdala by the infralimbic region of the medial prefrontal cortex, which results in the reduced fear response.(Caliskan and Albrecht, 2013) VNS increases corticotropin-releasing-hormone (CRH) expression in the hypothalamus (Hosoi et al., 2000) and CRH receptor 1 stimulation increases vagal modulation of heart rate variability (HRV).(Farrokhi et al., 2007; Porges, 2009, 1995) This reciprocal CRH - vagus nerve circuitry provides an important diagnostic and therapeutic link between stress and the ANS which can be explored in the pregnant sheep model.(Frasch et al., 2018b; Wakefield et al., 2019)

In the last few years, novel non-invasive methods of VNS have been developed which will not require surgical cervical VNS implants, have minimal to no side effects, and are low-cost.(Clancy et al., 2014; Frangos et al., 2015; Lamb et al., 2017; Liu et al., 2013; Ylikoski et al., 2017) It is now possible to conceive that the methodology presented in this manuscript could be a foundation for the development of non-invasive treatments that will benefit human neonates in a variety of conditions, notably, sepsis.(P. Burns et al., 2020)

The presence of left-right asymmetry of the vagus nerve's effect on FHR control was revealed in the intrasurgically performed unilateral VNS prior to and after Vx. It is well known that the right vagus nerve innervates the sinoatrial node and the left vagus innervating the atrioventricular node. Studies in the dog show that the right vagus produces a greater slowing in heart rate when compared to stimulation of the left vagus.(Randall and Ardell, 1985) This is the reason why the VNS probes used for epilepsy control are commonly installed on the left side to minimize the cardiac side effects on rhythm.(Ben-Menachem, 2002) Our findings suggest that this asymmetry is present, at least to a degree, as early as in a mature fetus near-term. Further studies are needed to quantify the degree of such maturity.

Various options of the presented generic approach are discussed elsewhere.(Burns et al., 2015) The VNS approach presented here is relatively easily performed in near-term fetal sheep owing to their large size. A specific feature of the presented approach is the use of the same electrodes to intermittently apply VNS pulse trains or to record VENG, as needed. The sampling rate of 10 kHz is probably at the lower end of the recommended settings and should be increased to 20 or even 30 kHz as reported by others in the meantime since the inception and performance of the presented studies.(Silverman et al., 2018; Steinberg et al., 2016; Zanos et al., 2018) Such a high sampling rate currently precludes wireless VENG data acquisition.

We observed some degree of neuronal degeneration due to VNS probes. However, we do not have data to support any functional deterioration over the period observed. This remains to be validated in future studies over the longer periods of time.

The physiological data are presented for the period of intrasurgical recordings and the post-surgical recovery during 72 hours. In our experiments, the approach proved to be stable for an additional 72 hours. The objectives and results of the experimental design we chose to pursue will be reported elsewhere. At this point, we have no indication that a longer period of stimulation and recordings would be problematic. The precise upper limit of recording duration under conditions of Vx and VNS requires further studies.

A number of very promising future applications of the technique presented is derived from the ever-growing number of sheep-specific molecular biology reagents and recent sheep genome sequencing.(Jiang et al., 2014) These recent developments have further promoted this animal model to be a very promising and powerful approach to understand healthy and pathological human fetal development on various scales of organization, from (epi)genome to integrative physiology.(Jiang et al., 2014), (Begum et al., 2012; Byrne et al., 2014; Lie et al., 2014; Nicholas et al., 2013; Wang et al., 2015; Zhang et al., 2010).

**Disclosures:**
No disclosures have been made.

**Acknowledgments:**
The authors gratefully acknowledge funding support from the Molly Towell Perinatal Research Foundation, Canadian Institutes of Health Research (CIHR), and Fonds de Recherche du Québec – Santé (FRQS) (to MGF) and CIHR-Quebec Training Network in Perinatal Research (QTNPR) (to LDD).
The authors wish to thank Esther Simard, Marco Bosa, Carl Bernard, Carmen Movila and Robin Pfaff for technical assistance. We thank Jan Hamanishi for the skillful graphical design assistance.

# References

Andersson, U., Tracey, K.J., 2012. Reflex principles of immunological homeostasis. Annu. Rev. Immunol. 30, 313–335. https://doi.org/10.1146/annurev-immunol-020711-075015

Ben-Menachem, E., 2002. Vagus-nerve stimulation for the treatment of epilepsy. Lancet Neurol. 1, 477–482. https://doi.org/10.1016/s1474-4422(02)00220-x

Borovikova, L.V., Ivanova, S., Zhang, M., Yang, H., Botchkina, G.I., Watkins, L.R., Wang, H., Abumrad, N., Eaton, J.W., Tracey, K.J., 2000. Vagus nerve stimulation attenuates the systemic inflammatory response to endotoxin. Nature 405, 458–462. https://doi.org/10.1038/35013070

Burns, P., Herry, C.L., Jean, K.J., Frank, Y., Wakefield, C., Cao, M., Desrochers, A., Fecteau, G., Last, M., Faure, C., Frasch, M.G., 2020. The neonatal sepsis is diminished by cervical vagus nerve stimulation and tracked non-invasively by ECG: a preliminary report in the piglet model. arXiv [q-bio.TO].

Burns, P., Liu, H.L., Kuthiala, S., Fecteau, G., Desrochers, A., Durosier, L.D., Cao, M., Frasch, M.G., 2015. Instrumentation of Near-term Fetal Sheep for Multivariate Chronic Non-anesthetized Recordings. J. Vis. Exp. e52581. https://doi.org/10.3791/52581

Burns, P.M., Castel, A., Javier, B., Cao, M., Paquet, M., Fecteau, G., Desrochers, A., Frasch, M., 2020. Cervical vagotomy and electroneurogram (VENG) approach in near-term fetal sheep. https://doi.org/10.6084/m9.figshare.7228307.v1

Caliskan, G., Albrecht, A., 2013. Noradrenergic interactions via autonomic nervous system: a promising target for extinction-based exposure therapy? J. Neurophysiol. 110, 2507–2510. https://doi.org/10.1152/jn.00502.2013

Carmel, E.N., Burns, P., Durosier, D., 2012. Fetal Brain MRI-Experiences in the Ovine Model of Cerebral Inflammatory Response (CIR). Reproductive.

Clancy, J.A., Mary, D.A., Witte, K.K., Greenwood, J.P., Deuchars, S.A., Deuchars, J., 2014. Non-invasive vagus nerve stimulation in healthy humans reduces sympathetic nerve activity. Brain Stimul. 7, 871–877. https://doi.org/10.1016/j.brs.2014.07.031

Cortes, M., Cao, M., Liu, H.L., Burns, P., Moore, C., Fecteau, G., Desrochers, A., Barreiro, L.B., Antel, J.P., Frasch, M.G., 2017. RNAseq profiling of primary microglia and astrocyte cultures in near-term ovine fetus: A glial in vivo-in vitro multi-hit paradigm in large mammalian brain. J. Neurosci. Methods 276, 23–32. https://doi.org/10.1016/j.jneumeth.2016.11.008

Durosier, L.D., Herry, C.L., Cortes, M., Cao, M., Burns, P., Desrochers, A., Fecteau, G., Seely, A.J.E., Frasch, M.G., 2015. Does heart rate variability reflect the systemic inflammatory response in a fetal sheep model of lipopolysaccharide-induced sepsis? Physiol. Meas. 36, 2089–2102. https://doi.org/10.1088/0967-3334/36/10/2089

Faltine-Gonzalez, D.Z., Layden, M.J., 2018. The origin and evolution of acetylcholine signaling through AchRs 2 in metazoans. bioRxiv. https://doi.org/10.1101/424804

Farrokhi, C.B., Tovote, P., Blanchard, R.J., Blanchard, D.C., Litvin, Y., Spiess, J., 2007. Cortagine: behavioral and autonomic function of the selective CRF receptor subtype 1 agonist. CNS Drug Rev. 13, 423–443. https://doi.org/10.1111/j.1527-3458.2007.00027.x

Frangos, E., Ellrich, J., Komisaruk, B.R., 2015. Non-invasive Access to the Vagus Nerve Central Projections via Electrical Stimulation of the External Ear: fMRI Evidence in Humans. Brain Stimul. 8, 624–636. https://doi.org/10.1016/j.brs.2014.11.018

Frasch, M.G., Burns, P., Benito, J., Cortes, M., Cao, M., Fecteau, G., Desrochers, A., 2018a. Sculpting the Sculptors: Methods for Studying the Fetal Cholinergic Signaling on Systems

and Cellular Scales. Methods Mol. Biol. 1781, 341–352. https://doi.org/10.1007/978-1-4939-7828-1_18

Frasch, M.G., Lobmaier, S.M., Stampalija, T., Desplats, P., Pallarés, M.E., Pastor, V., Brocco, M.A., Wu, H.-T., Schulkin, J., Herry, C.L., Seely, A.J.E., Metz, G.A.S., Louzoun, Y., Antonelli, M.C., 2018b. Non-invasive biomarkers of fetal brain development reflecting prenatal stress: An integrative multi-scale multi-species perspective on data collection and analysis. Neurosci. Biobehav. Rev. https://doi.org/10.1016/j.neubiorev.2018.05.026

Frasch, M.G., Mueller, T., Hoyer, D., Weiss, C., Schubert, H., Schwab, M., 2009. Nonlinear properties of vagal and sympathetic modulations of heart rate variability in ovine fetus near term. American journal of physiology.Regulatory, integrative and comparative physiology 296, R702–R707. https://doi.org/10.1152/ajpregu.90474.2008

Frasch, M.G., Walter, B., Friedrich, H., Hoyer, D., Eiselt, M., Bauer, R., 2007. Detecting the signature of reticulothalamocortical communication in cerebrocortical electrical activity. Clin. Neurophysiol. 118, 1969–1979. https://doi.org/10.1016/j.clinph.2007.05.071

Frasch, M.G., Zwiener, U., Hoyer, D., Eiselt, M., 2007. Autonomic organization of respirocardial function in healthy human neonates in quiet and active sleep. Early Hum. Dev. 83, Aug 25; [Epub ahead of print]. https://doi.org/10.1016/j.earlhumdev.2006.05.023

George, M.S., Ward, H.E., Jr, Ninan, P.T., Pollack, M., Nahas, Z., Anderson, B., Kose, S., Howland, R.H., Goodman, W.K., Ballenger, J.C., 2008. A pilot study of vagus nerve stimulation (VNS) for treatment-resistant anxiety disorders. Brain Stimul. 1, 112–121. https://doi.org/10.1016/j.brs.2008.02.001

Gotsch, F., Romero, R., Kusanovic, J.P., Mazaki-Tovi, S., Pineles, B.L., Erez, O., Espinoza, J., Hassan, S.S., 2007. The fetal inflammatory response syndrome. Clin. Obstet. Gynecol. 50, 652–683. https://doi.org/10.1097/GRF.0b013e31811ebef6

Herry, C.L., Burns, P., Desrochers, A., Fecteau, G., Durosier, L.D., Cao, M., Seely, A.J.E., Frasch, M.G., 2019. Vagal contributions to fetal heart rate variability: an omics approach. Physiol. Meas. https://doi.org/10.1088/1361-6579/ab21ae

Herry, C.L., Cortes, M., Wu, H.-T., Durosier, L.D., Cao, M., Burns, P., Desrochers, A., Fecteau, G., Seely, A.J.E., Frasch, M.G., 2016. Temporal Patterns in Sheep Fetal Heart Rate Variability Correlate to Systemic Cytokine Inflammatory Response: A Methodological Exploration of Monitoring Potential Using Complex Signals Bioinformatics. PLoS One 11, e0153515. https://doi.org/10.1371/journal.pone.0153515

Hosoi, T., Okuma, Y., Nomura, Y., 2000. Electrical stimulation of afferent vagus nerve induces IL-1beta expression in the brain and activates HPA axis. American journal of physiology.Regulatory, integrative and comparative physiology 279, R141–7.

Jiang, Y., Xie, M., Chen, W., Talbot, R., Maddox, J.F., Faraut, T., Wu, C., Muzny, D.M., Li, Y., Zhang, W., Stanton, J.A., Brauning, R., Barris, W.C., Hourlier, T., Aken, B.L., Searle, S.M., Adelson, D.L., Bian, C., Cam, G.R., Chen, Y., Cheng, S., DeSilva, U., Dixen, K., Dong, Y., Fan, G., Franklin, I.R., Fu, S., Fuentes-Utrilla, P., Guan, R., Highland, M.A., Holder, M.E., Huang, G., Ingham, A.B., Jhangiani, S.N., Kalra, D., Kovar, C.L., Lee, S.L., Liu, W., Liu, X., Lu, C., Lv, T., Mathew, T., McWilliam, S., Menzies, M., Pan, S., Robelin, D., Servin, B., Townley, D., Wang, W., Wei, B., White, S.N., Yang, X., Ye, C., Yue, Y., Zeng, P., Zhou, Q., Hansen, J.B., Kristiansen, K., Gibbs, R.A., Flicek, P., Warkup, C.C., Jones, H.E., Oddy, V.H., Nicholas, F.W., McEwan, J.C., Kijas, J.W., Wang, J., Worley, K.C., Archibald, A.L., Cockett, N., Xu, X., Wang, W., Dalrymple, B.P., 2014. The sheep genome illuminates biology of the rumen and lipid metabolism. Science 344, 1168–1173. https://doi.org/10.1126/science.1252806

Kwan, H., Garzoni, L., Liu, H.L., Cao, M., Desrochers, A., Fecteau, G., Burns, P., Frasch, M.G.,

2016. Vagus Nerve Stimulation for Treatment of Inflammation: Systematic Review of Animal Models and Clinical Studies. Bioelectron Med 3, 1–6.

Lamb, D.G., Porges, E.C., Lewis, G.F., Williamson, J.B., 2017. Non-invasive Vagal Nerve Stimulation Effects on Hyperarousal and Autonomic State in Patients with Posttraumatic Stress Disorder and History of Mild Traumatic Brain Injury: Preliminary Evidence. Front. Med. 4, 124. https://doi.org/10.3389/fmed.2017.00124

Liu, H.L., Garzoni, L., Herry, C., Durosier, L.D., Cao, M., Burns, P., Fecteau, G., Desrochers, A., Patey, N., Seely, A.J.E., Faure, C., Frasch, M.G., 2016. Can Monitoring Fetal Intestinal Inflammation Using Heart Rate Variability Analysis Signal Incipient Necrotizing Enterocolitis of the Neonate? Pediatr. Crit. Care Med. 17, e165–76. https://doi.org/10.1097/PCC.0000000000000643

Liu, R.P., Fang, J.L., Rong, P.J., Zhao, Y., Meng, H., Ben, H., Li, L., Huang, Z.X., Li, X., Ma, Y.G., Zhu, B., 2013. Effects of electroacupuncture at auricular concha region on the depressive status of unpredictable chronic mild stress rat models. Evid. Based. Complement. Alternat. Med. 2013, 789674. https://doi.org/10.1155/2013/789674

Lizier, J.T., 2014. JIDT: An Information-Theoretic Toolkit for Studying the Dynamics of Complex Systems. Frontiers in Robotics and AI 1, 11. https://doi.org/10.3389/frobt.2014.00011

Morris, G.L., 3rd, Mueller, W.M., 1999. Long-term treatment with vagus nerve stimulation in patients with refractory epilepsy. The Vagus Nerve Stimulation Study Group E01-E05. Neurology 53, 1731–1735. https://doi.org/10.1212/WNL.53.8.1731

Morrison, J.L., Berry, M.J., Botting, K.J., Darby, J.R.T., Frasch, M.G., Gatford, K.L., Giussani, D.A., Gray, C.L., Harding, R., Herrera, E.A., Kemp, M.W., Lock, M.C., McMillen, I.C., Moss, T.J., Musk, G.C., Oliver, M.H., Regnault, T.R.H., Roberts, C.T., Soo, J.Y., Tellam, R.L., 2018. Improving pregnancy outcomes in humans through studies in sheep. Am. J. Physiol. Regul. Integr. Comp. Physiol. https://doi.org/10.1152/ajpregu.00391.2017

Nitsos, I., Rees, S.M., Duncan, J., Kramer, B.W., Harding, R., Newnham, J.P., Moss, T.J., 2006. Chronic exposure to intra-amniotic lipopolysaccharide affects the ovine fetal brain. J. Soc. Gynecol. Investig. 13, 239–247. https://doi.org/10.1016/j.jsgi.2006.02.011

O'Keane, V., Dinan, T.G., Scott, L., Corcoran, C., 2005. Changes in hypothalamic-pituitary-adrenal axis measures after vagus nerve stimulation therapy in chronic depression. Biol. Psychiatry 58, 963–968. https://doi.org/10.1016/j.biopsych.2005.04.049

Patel, Y.A., Butera, R.J., 2015. Differential fiber-specific block of nerve conduction in mammalian peripheral nerves using kilohertz electrical stimulation. J. Neurophysiol. 113, 3923–3929. https://doi.org/10.1152/jn.00529.2014

Pavlov, V.A., Tracey, K.J., 2017. Neural regulation of immunity: molecular mechanisms and clinical translation. Nat. Neurosci. 20, 156–166. https://doi.org/10.1038/nn.4477

Pavlov, V.A., Tracey, K.J., 2015. Neural circuitry and immunity. Immunol. Res. 63, 38–57. https://doi.org/10.1007/s12026-015-8718-1

Pavlov, V.A., Tracey, K.J., 2012. The vagus nerve and the inflammatory reflex--linking immunity and metabolism. Nat. Rev. Endocrinol. 8, 743–754. https://doi.org/10.1038/nrendo.2012.189

Pena, D.F., Childs, J.E., Willett, S., Vital, A., McIntyre, C.K., Kroener, S., 2014. Vagus nerve stimulation enhances extinction of conditioned fear and modulates plasticity in the pathway from the ventromedial prefrontal cortex to the amygdala. Front. Behav. Neurosci. 8, 327. https://doi.org/10.3389/fnbeh.2014.00327

Porges, S.W., 2009. The polyvagal theory: new insights into adaptive reactions of the autonomic nervous system. Cleve. Clin. J. Med. 76 Suppl 2, S86–90.

    https://doi.org/10.3949/ccjm.76.s2.17

Porges, S.W., 1995. Cardiac vagal tone: a physiological index of stress. Neurosci. Biobehav. Rev. 19, 225–233.

Randall, W.C., Ardell, J.L., 1985. Differential innervation of the heart, in: Sipes, D., Jalife, J. (Eds.), Cardiac Electrophysiology and Arrhythmias. Grune and Stratton, New York, pp. 137–144.

Rozman, J., Ribaric, S., 2007. Selective recording of electroneurograms from the left vagus nerve of a dog during stimulation of cardiovascular or respiratory systems. Chin. J. Physiol. 50, 240–250.

Silverman, H.A., Stiegler, A., Tsaava, T., Newman, J., Steinberg, B.E., Masi, E.B., Robbiati, S., Bouton, C., Huerta, P.T., Chavan, S.S., Tracey, K.J., 2018. Standardization of methods to record Vagus nerve activity in mice. Bioelectronic Medicine 4, 3. https://doi.org/10.1186/s42234-018-0002-y

Steinberg, B.E., Silverman, H.A., Robbiati, S., Gunasekaran, M.K., Tsaava, T., Battinelli, E., Stiegler, A., Bouton, C.E., Chavan, S.S., Tracey, K.J., Huerta, P.T., 2016. Cytokine-specific Neurograms in the Sensory Vagus Nerve. Bioelectron Med 3, 7–17.

Svedin, P., Kjellmer, I., Welin, A.K., Blad, S., Mallard, C., 2005. Maturational effects of lipopolysaccharide on white-matter injury in fetal sheep. J. Child Neurol. 20, 960–964. https://doi.org/10.1177/08830738050200120501

Wakefield, C., Janoschek, B., Frank, Y., Karp, F., Reyes, N., Schulkin, J., Frasch, M.G., 2019. Chronic stress may disrupt covariant fluctuations of vitamin D and cortisol plasma levels in pregnant sheep during the last trimester: a preliminary report. arXiv [q-bio.TO].

Yan, E., Castillo-Melendez, M., Nicholls, T., Hirst, J., Walker, D., 2004. Cerebrovascular responses in the fetal sheep brain to low-dose endotoxin. Pediatr. Res. 55, 855–863. https://doi.org/10.1203/01.PDR.0000115681.95957.D4

Ylikoski, J., Lehtimaki, J., Pirvola, U., Makitie, A., Aarnisalo, A., Hyvarinen, P., Ylikoski, M., 2017. Non-invasive vagus nerve stimulation reduces sympathetic preponderance in patients with tinnitus. Acta Otolaryngol. 1–9. https://doi.org/10.1080/00016489.2016.1269197

Zanos, T., Silverman, H., Levy, T., Tsaava, T., Battinelli, E., Lorraine, P., Ashe, J., Chavan, S.S., Bouton, C., Tracey, K.J., 2018. Identification of cytokine-specific sensory neural signals in murine vagus nerve activity recordings. The Journal of Immunology 200, 43.12–43.12.

# Tables

**Table 1. Neuropathological findings in the vagus nerve subjected to the VNS probe.**

| Group | Side | Lesion | Degeneration: distal | Degeneration: proximal | Inflammation, edema: proximal | VNS probe |
|---|---|---|---|---|---|---|
| Twin ctrl | N/A | A | | | | No |
| Twin ctrl | N/A | N | | | | No |
| Twin ctrl | N/A | N | | | | No |
| Twin ctrl | N/A | N | | | | No |
| Twin ctrl | Left | N | | | | No |
| Twin ctrl | Right | N | | | | No |
| Vx+LPS800 | Left | N | | | | No |
| Vx+LPS800 | Right | A | | | | No |
| Vx+LPS800 | Left | A | 2 | | | No |
| Vx+LPS800 | Right | A | 1 | | | No |
| Vx+LPS800 | Left | N | | | | No |
| Vx+LPS800 | Right | N | | | | No |
| Efferent VNS | Left | A | 3 | | | Yes |
| Efferent VNS | Right | A | 4 | | 2 | Yes |
| Efferent VNS | Left | A | 3 | | 1 | Yes |
| Efferent VNS | Right | A | | 3 | 1 | Yes |
| Efferent VNS | Left | A | 3 | | 3 | Yes |
| Efferent VNS | Right | A | 4 | | 1 | Yes |
| Afferent VNS | Left | A | 2 | | 1 | Yes |
| Afferent VNS | Right | A | 4 | | 1 | Yes |
| Afferent VNS | Left | A | 4 | 1 | 3 | Yes |
| Afferent VNS | Right | A | 2 | 1 | 1 | Yes |
| Afferent VNS | Left | A | 3 | 1 | 3 | Yes |
| Afferent VNS | Right | A | 2 | | 2 | Yes |
| Afferent VNS | Left | A** | 1 | | 2 | Yes |
| Afferent VNS | Right | A | 2 | | 2 | Yes |
| Afferent VNS | Left | A | 2 | | 1 | Yes |
| Afferent VNS | Right | A | 2 | | 1 | Yes |
| Afferent VNS | Left | A | 3 | 2 | 1 | Yes |
| Afferent VNS | Right | A | 2 | | 1 | Yes |
| Afferent VNS | Left | A | | 1 | 2 | Yes |
| Afferent VNS | Right | A | 3 | 1 | | Yes |

The lesions were studied and quantified on a scale from 0 to 4, 0 being absent and 4 being > 51% of the specimen. Grade 1 corresponded to modest (-10%), Grade 2= mild(11-30%), Grade 3= moderate(31-50%), Grade 4= severe(+51%) lesion of the respective kind as listed below:

- Degenerative changes (Wallerian degeneration, ischemic changes)(affecting the distal half segment)
- Degenerative changes (Wallerian degeneration, ischemic changes)(affecting the proximal half segment)
- Multifocal granulocytic inflammation with interstitial edema and hemorrhages, proximal segment